\documentclass[multphys,vecphys]{svmult}
\def\Z{\;\>Z\!\!\!\!\!\! Z\;\;}

\def\thru#1{\mathrel{\mathop{#1\!\!\!/}}}
\def\1{\;1\!\!\!\! 1\;}
\def\Int#1#2{\int\!d^{#1}{#2}\,}
\def\R{\;\,R\!\!\!\!\!\!\!\> R\,\,}

\def\half{\hbox{${1\over 2}$}}

\def\lsim{\mathrel{\rlap{\lower4pt\hbox{\hskip1pt$\sim$}}
    \raise1pt\hbox{$<$}}}         
\def\gsim{\mathrel{\rlap{\lower4pt\hbox{\hskip1pt$\sim$}}
    \raise1pt\hbox{$>$}}}         

\def\tr{\,{\hbox{\rm tr}}\,}
\def\epm#1#2{\hbox{${\lower1pt\hbox{$\scriptstyle +#1$}}
\atop {\raise1pt\hbox{$\scriptstyle -#2$}}$}}

\newcommand     \bea     {\begin{eqnarray}} 
\newcommand     \eea     {\end{eqnarray}} 
\newcommand     \beq     {\begin{equation}} 
\newcommand     \eeq    {\end{equation}} 

\usepackage{makeidx}         
\usepackage{graphicx}        
\usepackage{multicol}        
\usepackage[bottom]{footmisc}

\makeindex             

\begin{document}
\pagestyle{empty}
{\large
\begin{center}
\vspace*{0.5cm}
{\huge \bf Spin in Quantum Field Theory}\\\vspace*{1cm}
{\Large Stefano Forte}\\\medskip
\vspace{0.6cm}  {\it
Dipartimento di Fisica, Universit\`a di Milano and\\
INFN, Sezione di Milano,\\
Via Celoria 16, I-20133 Milano, Italy}\\
\vspace*{2.5cm}
{ \bf Abstract}
\end{center}

{ I introduce spin in field theory 
by emphasizing the close connection between quantum field
theory and quantum mechanics. First, I show that  the spin--statistics
connection can be derived in quantum mechanics without relativity or
field theory. Then, I
discuss path integrals for spin without using spinors. 
Finally, I show how spin can be quantized
in a path--integral approach,
without introducing anticommuting variables.}

\vspace*{1.cm}

\begin{center}
 Invited lectures at the\\ 
43 Internationale Universit\"atswochen f\"ur Theoretische Physik
\\
Schladming, Austria, February - March  2005\\
{\it to be published in the proceedings}
\end{center}
\vfill
IFUM-844/FT\hfill
July 2005
}
\eject

\pagestyle{plain}

\title*{Spin in Quantum  Field Theory}
\author{Stefano Forte}
\institute{Dipartimento di Fisica, Universit\`a di Milano and\\ INFN,
  Sezione di Milano\\ via Celoria 16, I-10129 Milano, Italy;\\
\texttt{forte@mi.infn.it}
}
%
%
\maketitle

I introduce spin in field theory 
by emphasizing the close connection between quantum field
theory and quantum mechanics. First, I show that  the spin--statistics
connection can be derived in quantum mechanics without relativity or
field theory. Then, I
discuss path integrals for spin without using spinors. 
Finally, I show how spin can be quantized
in a path--integral approach,
without introducing anticommuting variables.

\section{From Quantum Mechanics to Field Theory}
\label{sec:intro}
Even though everybody learns about spin in their childhood in the
context of nonrelativistic quantum mechanics, many of the more
interesting dynamical features of spin are only introduced in relativistc
quantum field theory. In these lectures, which were originally
addressed to an audience of (mostly) condensed-matter physicists, I
discuss some relevant aspects of spin dynamics in quantum field
theory by showing their origin in quantum mechanics. In the first
lecture, after a brief discussion of the way spin appears in
nonrelativistic (Galilei invariant) or relativistic (Lorentz
invariant) dynamics, I show how the spin--statistics connection can be
obtained with minimal assumptions in nonrelativistic quantum
mechanics, without invoking relativity or field theory.
In the second lecture I show how spin can be quantized in a
path--integral approach with no need for introducing quantum
fields. In the third lecture I discuss the dynamics of
relativistic spinning particles and show that its quantization can be
described without using anticommuting variables. A fourth lecture was
devoted to the quantum breaking of chiral symmetry -- the
axial anomaly -- and its origin in the structure of the spectrum of the Dirac
operator, but since this subject is already covered in many classic lectures~\cite{jackiw} it
will not be covered here. 
 We will see that even though the standard methods of quantum
field theory are much more practical for actual calculations,
a purely quantum--mechanical approach helps in understanding the
meaning of field--theoretic concepts.

\section{Spin and Statistics}
\label{sec:spinstat}

\subsection{The Galilei Group and the Lorentz Group}
\label{sec:galo}

In both relativistic and non-relativistic dynamics we can understand
the meaning of quantum numbers in terms of the  symmetries
of the
Hamiltonian and the Lagrangian and associated action. Indeed, 
the invariance of the Hamiltonian determines the 
spectrum of physical
states: eigenstates of the Hamiltonian  are classified by the
eigenvalues of operators which commute with it, and this gives the set
of observables which are conserved by time evolution. However, the
invariance of the dynamics is defined by the  invariance
of the action. This is bigger than that of the Hamiltonian, because it also
involves time--dependent transformations. For example, in a
nonrelativistic theory the action must be invariant under Galilei
boost: the change between two frames that move at constant velocity
with respect to each other. But the Hamitonian in general doesn't possess
this invariance: Galilei boosts obviously change the values of the
momenta, and the Hamiltonian in general depends on them. 
 The set of operators which commute with all transformations
that leave the action invariant defines the quantum numbers carried by
elementary excitations of the system (elementary particles).

A nonrelativistic theory must have an action which is invaraint upon
the Galilei group. The Galilei transformations, along with the
associate quantum-mechanical operators are~\cite{sym}:
\begin{itemize}
\item { space translations}: $x_i\to x_i^\prime=x_i+a_i$;\hfill  $P_i=-i\partial_i$
\item { time translation}: $t\to t^\prime=t+a$;\hfill 
  $H=i \frac {d}{dt}$ 
\item { Galilei boosts}: $x_i\to x_i^\prime=x_i+ v_i t;\>
  p_i\to p_i+m v_i$;\hfill 
$K_i=-it\partial_i-mx_i$
\item { rotations}: $x_i\to x_i^\prime=R_{ij} x_j$;\hfill  $J_i=\epsilon_{ijk}x^j \partial_k+{ \sigma_i}$
\end{itemize}
The generator of rotations is the sum of orbital angular momentum and {\it spin}.

The generators of the Galilei group form the Galilei algebra:
\bea
&&[J_i,J_j]=\epsilon_{ijk}J_k;\quad [P_i,P_j]=0; \quad
[K_i,K_j]=0;\quad
[J_i,H]=[K_i,H]=0;\nonumber\\\label{galagebra}
&&[k_i,H]=i P_i;\quad
[J_i,P_j]=\epsilon_{ijk}P_k;\quad
[J_i,K_j]=\epsilon_{ijk}K_k;\quad [K_i,P_j]= i M \delta_{ij}.
\eea
In order to close the algebra it is necessary to introduce a
(trivial) mass operator $M$ which commutes with everything else:\\
\beq\label{malgebra}
[M,P_i]=[M,K_i]=[M,J_i]=[M,H]=0.
\eeq

The Casimir operators, which commute with all generators, are
\beq\label{galcasop}
C_1=M;\quad C_2=2 MP_0- P_iP_i;\quad C_3=\left(M
J_i-\epsilon_{ijk}P_jK_k\right)\left(M
J_i-\epsilon_{ilm}P_lK_m\right).
\eeq
In terms of quantum-mechanical operators the Casimirs correspond to
\begin{itemize}
\item $C_1=m$\quad{ (mass)};
\item $\frac{1}{2m} C_2=-i\frac{d}{dt}-\frac{p^2}{2m}$\quad{ (internal
  energy)}; 
\item $C_3=\sigma_i\sigma_i$\quad{ ({\bf spin})}.
\end{itemize}
We see that spin is one of the three numbers which classify
nonrelativistic elementary excitations, along with mass and internal
energy.

In the relativistic case, the action is invariant under the 
Poincar\'e group. The transformations and associate operators are now:
\begin{itemize}
\item { translations}: $x_\mu\to x_\mu^\prime=x_\mu+a_\mu$;\hfill
  $P_\mu=-i\partial_\mu$ 
\item { Lorentz transf.}:\\ 
$x^\mu\to {x^\prime}^\mu
  =\Lambda^\mu{}_\nu x^\nu;\>
p^\mu\to {p^\prime}^\mu
  =\Lambda^\mu{}_\nu p^\nu$;\\
\phantom{.}\hfill 
$J^{\mu\nu}=x^\mu P^\nu-x^\nu P^\mu+\Sigma^{\mu\nu}$\\
\phantom{.}\hfill{ rotations}:
  $J_i=\frac{1}{2}\epsilon_{ijk}J^k=\epsilon_{ijk}x_jP_k+{
  \sigma^i}$\\\phantom{.}\hfill{ boosts}:
  $K_i=J^{i0}$.\end{itemize}

The Poincar\'e generators form the algebra
\bea
&&[J^{\mu\nu},J^{\rho\sigma}]=i\left(g^{\mu\rho}J^{\nu\sigma}-
g^{\nu\rho}J^{\mu\sigma}+
g^{\mu\sigma}J^{\nu\rho}+
g^{\nu\sigma}J^{\mu\rho}
\right);\nonumber\\ \label{poinalgebra}
&&[P^\mu,J^{\rho\sigma}]=-i\left(g^{\mu\rho}P^\sigma-g^{\mu\sigma}P^\rho\right); \qquad
[K^\mu,P^\nu]=0\eea
Explicitly, the algebra of boosts and rotations is
\bea
[J_i,J_j]=\epsilon_{ijk}J_k;\quad [J_i,K_j]=\epsilon_{ijk}K_k;\quad
[K_i,K_j]={ -i\epsilon_{ijk}K_k}\nonumber\\\label{borot} 
[J_i,P_j]=\epsilon_{ijk}P_k;\quad[K_i,H]=i P_i;\quad[K_i,P_j]=
      { i H} \delta_{ij}.\eea

The Casimir operators are now just two:
\beq\label{poincasop}
 C_1=P_\mu P^\mu;\quad C_2=W_\mu W^\mu,
\eeq
in terms of the momentum generator and the {\it Pauli-Lubanski} operator
\beq
W^\mu=\epsilon^{\mu\nu\rho\sigma}P_\nu J_{\rho\sigma}.
\label{paluop}
\eeq
The corresponding quantum-mechanical operators are
\begin{itemize}
\item $C_1=P^2$; { eigenvalue} $M^2$~{ (mass)};\\
\item $C_2=W^2=m\sigma^2$;  
  {eigenvalue} $M^2s(s+1)$~{ (mass$\times${\bf  spin})},
\end{itemize}
\noindent where the latter identification is clear if one chooses the
rest frame, as we shall discuss in greater detail in section~\ref{sec:spinfer}.

Galilei transformations can be obtained from Poincar\'e
transformations in the limit
$v\ll 1$ by assuming the scaling laws  $M\sim 1$, $J\sim
1$, $P\sim v$, $H\sim v^2$  $K\sim 1/v$.

Summarizing, both in nonrelativistic and relativistic theories spin is
one of the quantum numbers that classify elementary excitations. 
In quantum mechanics, the
state vectors of physical systems are expanded on a basis of
irreducible representations of the rotation group (in the
nonrelativistic case) or the Lorentz group (in the relativistic
case).
In quantum field theory, one--particle states are, respectively,
Galilei or Poincar\'e irreducible representations. In the relativistic
case, rotations are implicitly defined
by the Pauli-Lubanski vector eq.~(\ref{paluop}) as the subgroup of the
Lorentz group which leaves the four-momentum invariant.
 
In more than two spatial dimensions the rotation group  O($d$) is doubly
connected
(i.e., $\pi_1[\hbox{O($d$
)}]=\Z_2$); its
universal cover is the group Spin($d$), which, in the usual $d=3$ case,
is isomorphic
to SU(2). When
$d=2$ (planar systems) the rotation group is O(2), which, being isomorphic to the circle $S^1$
is infinitely connected ($\pi_1[O(2)]=\Z$); its universal
cover is the real line $\R$.
It follows that in more than two dimension
the wave function can carry either a simple-valued (Bosons) or a 
double valued (Fermions)
representation of the rotation group, and
in two dimensions it may carry an arbitrarily multivalued one (anyons~\cite{forte}).

The multivaluedness of the 
representation  of rotations is is classified by the value of the
phase which the wave function acquires upon rotation by $2\pi$ about
an arbitrary axis (the $z$ axis, say):
\beq
R_z^{2\pi}\psi(q_1,\dots,q_n)=e^{2\pi iJ_z}\psi(q_1,\dots,q_n)=e^{2\pi
  i \sigma}\psi(q_1,\dots,q_n), \label{spindef}
\eeq
where $J_z=L_z+\sigma$, and in the last step we have used the fact
that the spectrum of orbital angular momentum is given by the
integers, so upon $2\pi$ rotation it is only spin that contributes to
the phase.
\subsection{Statistics and Topology}
\label{sec:statop}

The wave function for a system of $n$ identical particles must be
invariant in modulus, and thus acquire a phase, 
 upon interchange of the
full set of quantum  numbers $q_i$ of the $i$-th and $j$-th particle:
\beq
\psi(q_1,\dots, q_{ i},\dots,q_{ j},\dots,q_n)=e^{2\pi i\sigma}\psi(
q_1,\dots, q_{ j},\dots,q_{ i},\dots,q_n).
\label{statdef}
\eeq
The parameter $\sigma$, which is only defined modulo integers,
 is the statistics of particles $i$, $j$. We now
prove the spin-statistics theorem, which states that the
statistics is a universal property of particles $i$, $j$, and it is
equal to their spin (also in d=2, where the spin as we have seen can
be generic).

The proof is based on an analysis of the quantisation of systems
defined on topologically nontrivial configuration spaces. Indeed, if
${\cal C}_d$ is the configuration space for a single particle in $d$
dimensions, the
configuration space for a system of $n$ particles in $d$
dimensions is
\beq
\label{config}
 \bar{{\cal C}_d}^n={C_d}^{n}-{\cal D},
\eeq
where ${\cal D}$ is the set of points where the full set of quantum
numbers of two or more particles coincide. These points must be
excised from space, otherwise eq.~(\ref{statdef}) with $x_i=x_j$
implies that  necessarily $\sigma=0$ . 

If the particles are identical, points which differ by their
interchange must be identified. The configuration space then becomes
the coset space
\beq
\label{cospace}
{{\cal C}_d}^n={\bar{{\cal C}_d}^n\over S_n},
\eeq
where $S_n$ is the group of permutations of $n$ objects.
The topological structure of the configuration space changes when
going from two to more than two dimensions, just like the topological
structure of the rotation group discussed in section~\ref{sec:statop}.
Indeed, if $d=2$ the  space eq.~(\ref{config}), i.e. before dividing out
permutations, is multiply connected: a closed path traversed by 
the $i$-th particle in which particle $j$ is inside the loop formed by
particle $i$ cannot be deformed into a path in which
particle $j$ is outside the loop. The configuration space
  ${\cal C}_2^n$ is then also multiply connected, and its
fundamental group  is the braid group 
{ $\pi_1({\cal C}_2^n)=B_n$}, as we shall discuss explicitly below. 

In
more than two dimensions, the space 
$\bar{\cal C}_d^n$ is simply connected: all closed path traversed by a
particle can be continuously deformed into each other, because
in more than two dimensions one cannot distinguish the inside of a
one-dimensional curve from its
outside. However, the configuration space
${\cal
  C}_d^n$ is multiply connected. This implies that 
 a topologically nontrivial closed path in ${\cal
  C}_d^n$ must correspond to an open path in  $\bar{\cal
  C}_d^n$, because all closed paths in   $\bar{\cal
  C}_d^n$ can be deformed into each other. Furthermore, points in
 ${\cal
  C}_d^n$   that correspond to same point in $\bar{\cal
  C}_d^n$ are in one-to-one correspondence with elements of $S_n$, because 
  $S_n$ acts effectively, i.e. only the identity of $S_n$ maps all
  points of $\bar{\cal C}_d^n$ onto thenselves. It follows that
  equivalence classes of paths in  $\bar{\cal C}_d^n$ are in
  one-to-one correspondence with elements of the permutation group: 
\beq\label{topcs}
\pi_1({\cal C}_d^n)=S_n.
\eeq
Hence, the multiply connected nature of the configuration space is
directly linked with the presence of identical particle, and
specifically to the response of the system upon permutations, i.e. to
statistics. 

\begin{figure}
\centering
\includegraphics[width=0.65\linewidth]{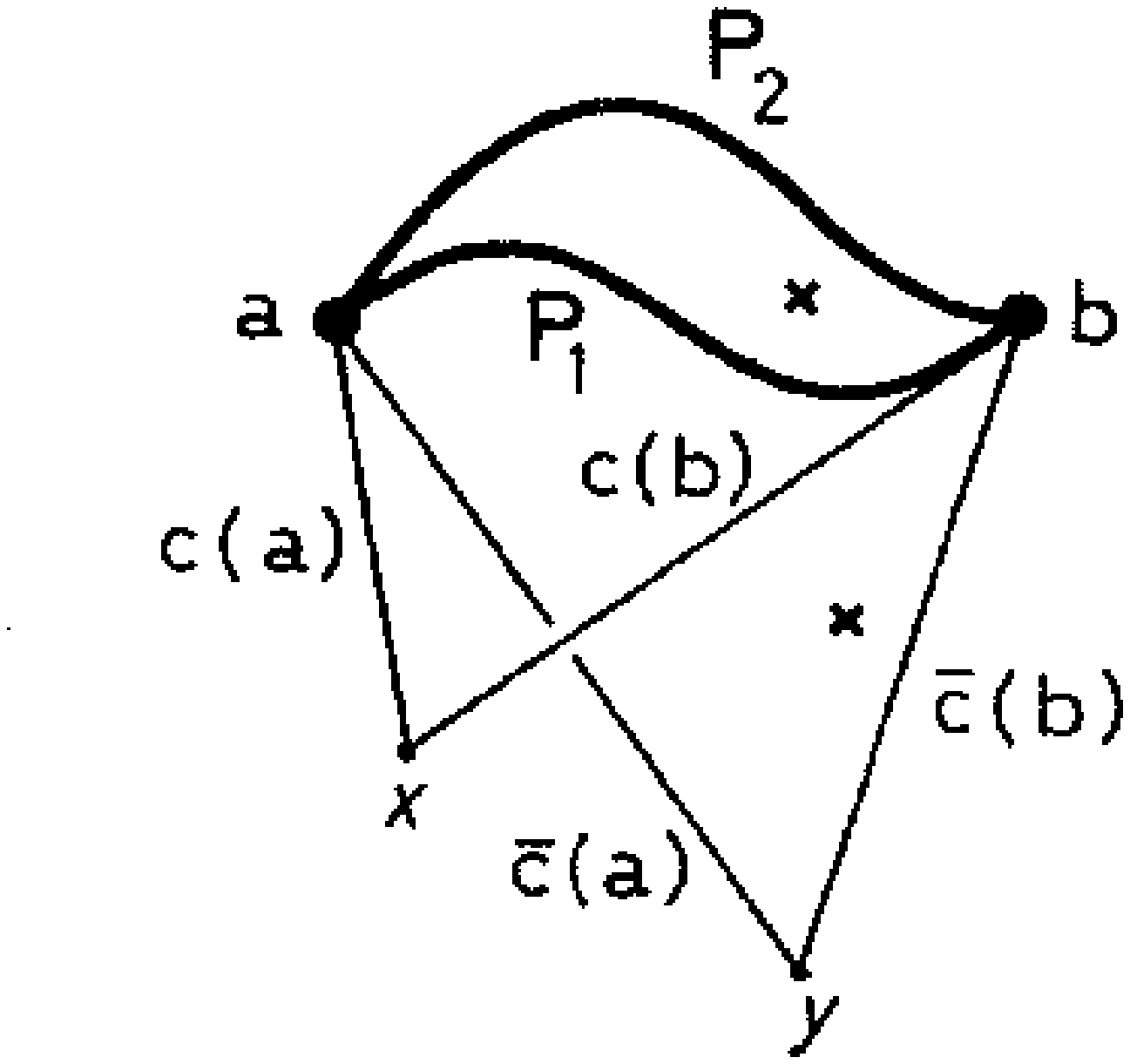}
\caption{Paths $P_i$ are assigned to homotopy classes by connecting
  them to a base point through a mesh. Changing the base point from
  $x$ to $y$ can change the absolute class assigment of a path, but
  not the relative assignment of a pair of paths.}
\label{fig:mesh}
\end{figure}
Therefore, let us consider quantization on a multiply connected
space. The way nontrivial statistics is obtained can be
understood by studying this problem in
a path--integral approach~\cite{laidlaw}, where transition amplitudes
are written in terms of the propagator $ K(q^\prime,q)$
\beq
\label{piamp}
S_{fi}\equiv\langle\psi_f|\psi_i\rangle=\langle \psi_f|q^\prime t^\prime\rangle\langle q^\prime t^\prime|
q t\rangle\langle q t | \psi_i\rangle
=\int\!dq\,dq^\prime\, \psi^*_f(q^\prime) K(q^\prime,q) \psi_i(q),\eeq
which in turn can be written as a sum over paths
\beq
\label{piprop}
{K(q^\prime,t^\prime;q,t)=\int_{q(t)=q;\>q(t^\prime)=q^\prime}\!\!\!\!
\!\!\!\!\!\!\!\!Dq(t_0)\>e^{i\int_t^{t^\prime}\!dt_0\,L[q(t_0)]}.}
\eeq

Closed paths on a multiply connected space fall into homotopy
classes. Moreover,  open paths can  also be
classified in homotopy classes by a choice of {\it mesh}
(figure~\ref{fig:mesh}). Namely, one chooses an arbitrary reference
point $x$ (base point) and then one assign to each point in space a
path connecting it to the base point. The homotopy class of an open
path can then be defined as the
homotopy class of the closed path formed by the given
open path and the mesh that connects it to the base point.
Once all paths (closed and open) are grouped into equivalence classes,
the path integral is in general  defined as follows
\beq
\label{mulconpi}
K(q^\prime,t^\prime;q,t)=\sum_\alpha \chi(\alpha)
K^\alpha(q^\prime,t^\prime;q,t),
\eeq
where  
$ K^\alpha(q^\prime,t^\prime;q,t)$ is computed including in the sum
over paths only paths in the $\alpha$-th homotopy class, and
$\chi(\alpha)$ are weights which depend only on the equivalence
class (homotopy class) of a given path. 

The weighted sum eq.~(\ref{mulconpi}) must satisfy the following
physical requirements:
\begin{itemize}
\item (a) physical result must be independent of the choice of mesh;
\item (b) amplitudes must satisfy the superposition principle, which
  in turn implies the convolutive property 
\beq
\label{conprop}
K(q{''},t{''};q,t)=\int d q^\prime \langle q{''} t{''}|
q^\prime t^\prime\rangle\langle q^\prime t^\prime|
q t\rangle=\int d q^\prime K(q{''},t{''}q^\prime t^\prime;)K(q^\prime
t^\prime;q,t).
\eeq
\end{itemize}
The  necessary and sufficient condition for these requirements to be
satisfied is that the weights $\chi(\alpha)$ satisfy
\bea
\label{wphase}
|\chi(\alpha)|&=&1\\
\label{wconv}
\chi(\alpha\circ\beta)&=&\chi(\alpha)\chi(\beta),\eea
where in eq.~(\ref{wconv}) $\alpha$ and $\beta$ are the homotopy
classes of paths with a common endpoint, and  $\alpha\circ\beta$ is
the homotopy class of the path
obtained by joining them.

\begin{figure}
\centering
\includegraphics[width=0.65\linewidth]{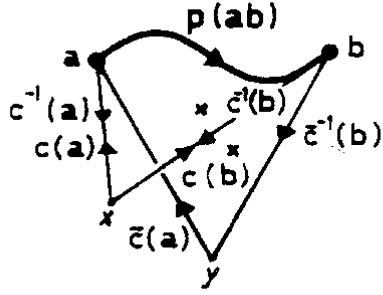}
\caption{Graphical representation of eq.~(\ref{mchange})}
\label{fig:mchange}
\end{figure}
The proof that eq.~(\ref{wconv}) implies property (b) is immediate:
\beq
\label{convsuff}
\sum_\gamma\chi(\gamma)K^\gamma(q^\prime,t^\prime;q,t)=\sum_{\alpha,\beta}\chi(\alpha)\chi(\beta)
\int d q^\prime K^\alpha(q{''},t{''}q^\prime t^\prime;)K^\beta(q^\prime
t^\prime;q,t).\eeq
The proof that eq.~(\ref{wphase}) implies property (a) is also
easy: let $P$ be the closed path obtained composing the
open path $p$ which connectes points $a$ and $b$ with a
mesh $C$ (figure~\ref{fig:mchange})).
Upon changing the mesh to $\bar C$, 
the path $P$ becomes the path $\bar P$, which
in turn can be obtained by composing $P$ with the closed paths
$\lambda\equiv \bar C(a)C^{-1}(a)$ and $\mu=C(b)\bar C^{-1}(b)$:
\bea
\label{mchange}
\bar P(ab)&=&\bar C(a)p(ab)\bar C^{-1}(b)\nonumber\\
 &=&\bar C(a)C^{-1}(a)C(a)p(ab)C^{-1}(b)C(b)\bar C^{-1}(b)\\
&=&\lambda P(ab)\mu.\nonumber
\eea
Because $\mu$ and $\lambda$ do not depend on the original path, but
 only on the two meshes, the factor $\chi(\lambda\mu) $ which relates
 the two class assignments 
\beq
\label{clrel}
\bar\chi(\alpha)=\chi(\lambda\mu) \chi(\alpha)
\eeq
is universal. It follows that 
\beq
\label{complaw}
\sum_\gamma \bar \chi(\gamma)
K^\alpha(q^\prime,t^\prime;q,t)=
\chi(\lambda\mu)\sum_\gamma  \chi(\gamma)
K^\alpha(q^\prime,t^\prime;q,t),
\eeq
so if $\chi$ are phases the transition probability is unchanged.
 
This proves that conditions (\ref{wphase}-\ref{wconv}) are sufficient
for requirements (a-b) to be satisfied, the proof that they are also
necessary is somewhat more technical and we shall omit it.
Conditions eq.~(\ref{wphase}-\ref{wconv}), taken jointly, 
mean that phases $\chi$ provide one-dimensional unitary representation of
$\pi_1({\cal C}_d^n)$, i.e. the permutation group $S_n$ (in more than
two dimensions) or the braid group (in two dimensions).

\subsection{Bosons, Fermions and Anyons}
\label{bofeany}

The relation between spin and statistics now follows from the
structure of the path integral. First, we observe that
there are only two unitary one-dimensional irreducible representations
of the
permutation group $S_n$: the trivial one (where $\chi=1$ 
for all permutations),  and the
alternating one, where $\chi=1$ if the permutation is even and
$\chi=-1$ if it is odd (i.e., if they may be performed by an even or
odd number of interchanges, respectively).
Now, note that the
 wave function at time { $t$} is given by the path integral in terms of
  some boundary condition at time  $t_0$:
\beq
\label{timev}
\langle q,t|\psi\rangle=\int dq_0 K(q,t;q_0,t_0)\psi_0(q_0,t_0).
\eeq
Two evolutions that lead to final states
which only differ by the interchange of the coordinates $q_i$, $q_j$
in configuration space differ by the factor $\chi$: hence,
$\chi=-1$ correspond to $\sigma=\half$ ($\sigma=0$). However, an
interchange of coordinates $q_i$, $q_j$ can also be realized by a
rotation by $\pi$ of the system about any axis through the center of
mass of the two particles (or a rotation about any axis followed by a
translation), which in turn is generated by the corresponding angular
momentum operator
\beq
\label{rotint}
|q_j q_i\rangle =e^{i \pi J^{ij}_{z}}|q_i q_j\rangle,
\eeq
where $J^{ij}_z$ is the component along the (arbitrarily defined) $z$
axis of the angular momentum of particles $i$,~$j$. The constraint
that $\sigma$ can only be either integer or half-integer is understood
as a consequence of the trivial fact that two interchanges, or a
rotation by $2\pi$, must bring back to the starting configuration.

\begin{figure}
\centering
\includegraphics[width=.65\linewidth]{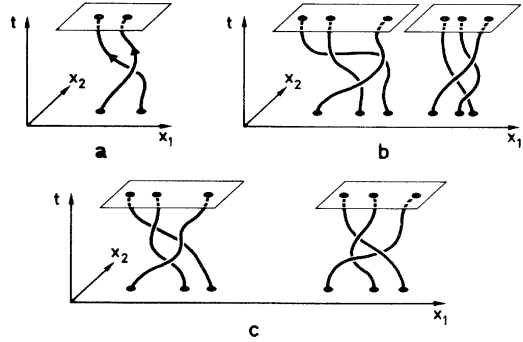}
\caption{Braids defined by particles' trajectories}
\label{fig:braids}
\end{figure}
It follows that
if $\chi=-1$, so $\sigma=\half$, and the
spectrum of
$J^{ij}_z$ is given by the odd integers. We can then view  the
contribution of $\chi$ to the path integral as the result of having
added an extra internal effective interaction, which shifts the
angular momentum of the pair of particles $i$,~$j$ by an integer,
i.e. the angular momentum of each particle by a half-integer. 
This establishes the spin--statistics theorem in a nonrelativistic
theory. The results is a consequence of the fact that fermionic
statistics, which is usually given as a property of wave functions,
has been lifted through the path--integral formalism  to a property of
particle paths, and attributed to a weight given to paths.
The fact that either trivial or alternating representations of
permutations are possible is then directly related to the existence of
either single--valued or double--valued representations of rotations.

In order to understand this better, let us now
 consider the case of planar systems~\cite{forte}, both
because we can then generalize this spin-statistics connection to
arbitrary spin and statistics (anyons), and also because we can then
work out an explicit representation for the effective interaction
associated to the $\chi$ weigths, which will lead us to the spin
action which we shall then discuss in the next section.
\begin{figure}
\centering
\includegraphics[width=.65\linewidth]{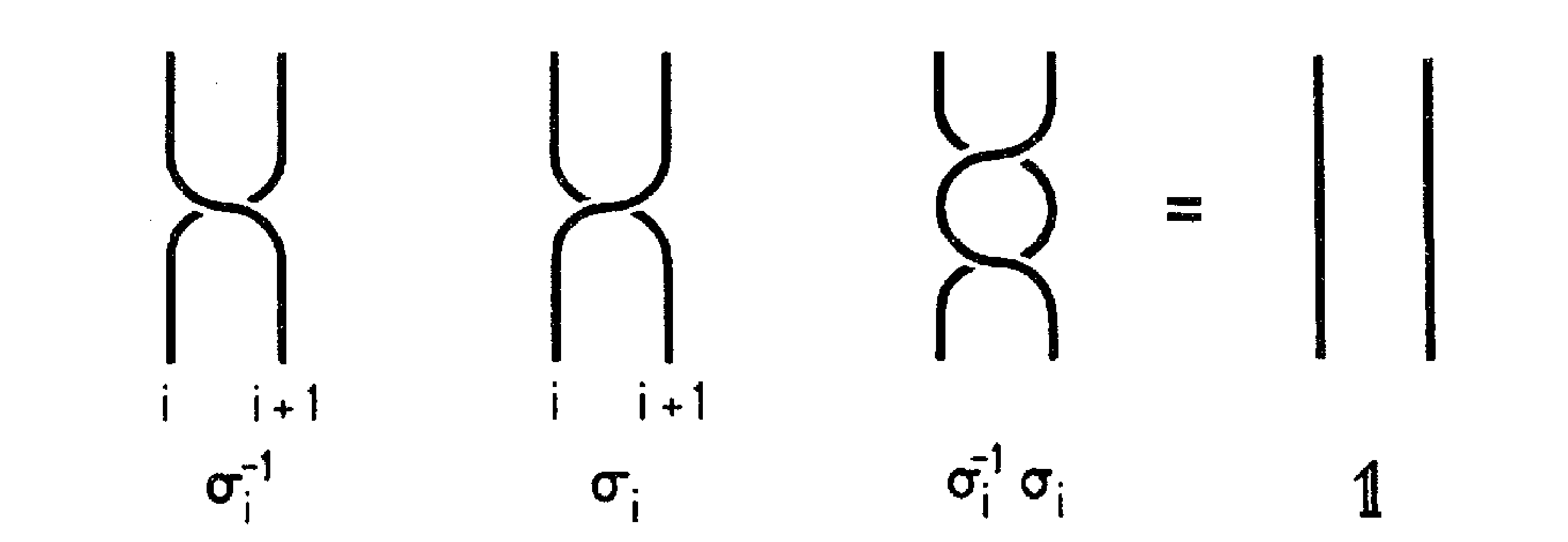}
\caption{The exchange operator $\sigma_i$ and its inverse\hfill}
\label{fig:sigop}
\end{figure}

\begin{figure}
\centering
\includegraphics[width=.65\linewidth]{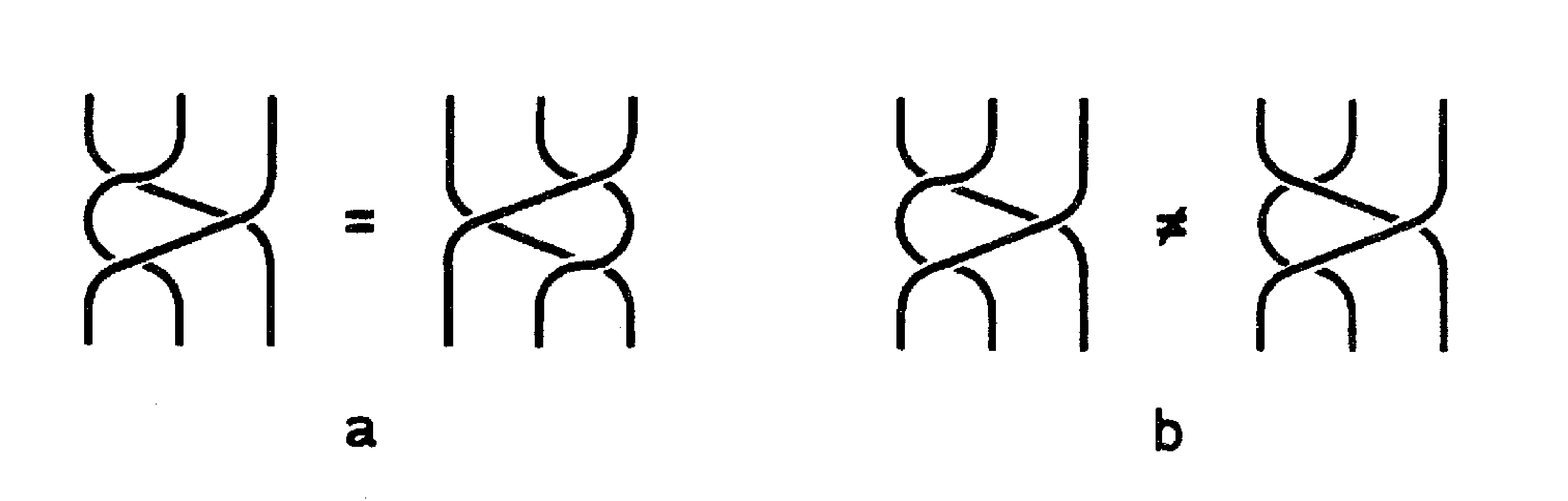}
\caption{Graphical representation of eq.~(\ref{eqbraids}) (a) and eq.~(\ref{neqbraids}) (b)}
\label{fig:enebr}
\end{figure}

In $d=2$, $\chi^\alpha$ provide an abelian irreducible representation of the braid
group. Indeed, each particle 
trajectory on a multiply--connected space defines
an inequivalent braid
(figure~\ref{fig:braids}). Each braid, in turn, is uniquely defined as
a sequence of interchanges of pairs of neighbouring particles. This
can be represented algebraically by introducing the operator
$\sigma_i$ which exchanges particles $i$ and $i+1$ (figure~\ref{fig:sigop}). 
Two braids are equivalent if they can be deformed into each
other. For instance (figure~\ref{fig:enebr})
\bea\label{eqbraids}
\sigma_i\sigma_{i+1}\sigma_i&=&\sigma_{i+1}\sigma_i\sigma_{i+1}\\
\label{neqbraids}
\sigma_i\sigma_{i+1}\sigma_i&\not=& \sigma_i\sigma_{i+1} \sigma_i^{-1}.
\eea

\begin{figure}
\centering
\includegraphics[width=.53\linewidth]{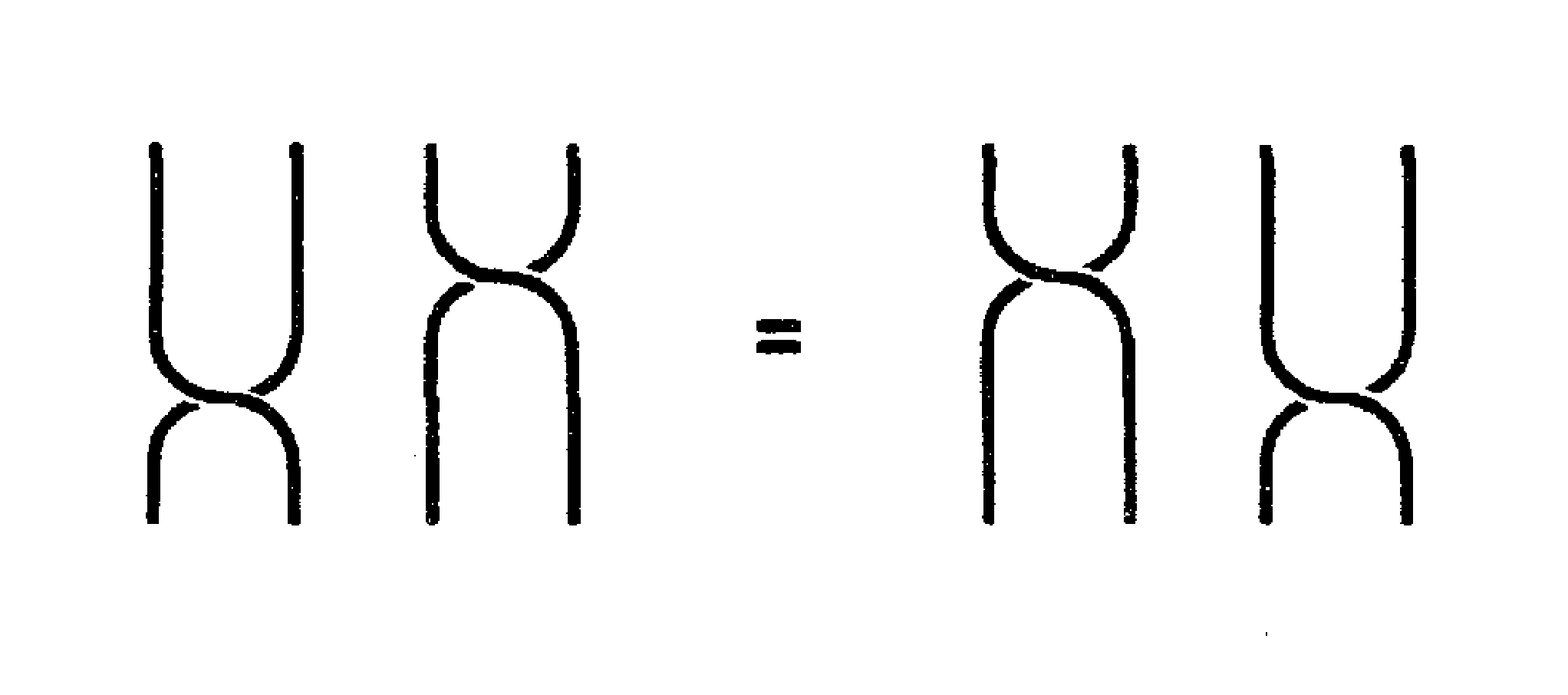}
\caption{Graphical representation of eq.~(\ref{indbraids})}
\label{fig:indbr}
\end{figure}
In fact, all independent relations between braids are
eq.~(\ref{eqbraids}) and
\beq
\label{indbraids}
\sigma_i\sigma_j=\sigma_j\sigma_i \quad\hbox{if}\> |i-j|>1.
\eeq
In terms 
of  $\chi$ eq.~(\ref{indbraids}) implies
\beq
\label{chiindbraids}
\chi(\sigma_i\sigma_j)=\chi(\sigma_i)\chi(\sigma_j) \quad \hbox{if}\>
  |i-j|>1,
\eeq
while eq.~(\ref{eqbraids})  implies
\beq
\label{chieqbraids}
\chi(\sigma_i)=\chi(\sigma_j)\quad \hbox{ for all}\> i,j.
\eeq

Equations~(\ref{chiindbraids},\ref{chieqbraids})
in turn imply that  the weight for a generic path (braid) is
\beq
\label{pweight}
\chi(\sigma_{i_1}\dots\sigma_{i_n})=\chi(\sigma_{i_1})\dots\chi(\sigma_{i_n})=
\exp\left(2i\sigma\sum_{k=1}^n \epsilon_k\right),
\eeq
where $\epsilon=+1$ for a direct exchange and  $\epsilon=-1$  for its
  inverse $\sigma_i^{-1}$, and $\sigma$ coincides with the statistics
  parameter eq.~(\ref{statdef}). The cases of bosons and fermions are
  recovered when $\sigma=0$ or $\sigma=\half$, respectively, but now
  $\sigma$ can take any real value (anyons). Indeed, in two dimensions two
  subsequent interchanges do not necessarily take back to the starting
  point, because a path where particle $i$ traverses a loop encircling
  particle $j$ cannot be shrunk to a point (identity). Hence, two
  interchanges do not necessarily bring back to the starting
  configuration, and the
  constraint that $2\sigma=1$ no longer applies. Accordingly, as
  already mentioned, in two dimension the rotation group admits
  arbitrarily multivalued representations.

The $\chi$ weights can be represented explicitly in 
in terms of the variation of relative polar angle $\Theta(\vec
x)\equiv\tan^{-1}\left(\frac{x^2}{x^1}\right)$ of particles $i$ and
$j$
along the particles' paths:
\beq
\label{chithrel}
\chi=\exp\left(-2 i
  \sigma\sum_{i<j}\Delta\Theta_{ij}\right)=
 \exp\left(-i
  \sigma\sum_{i\not=j}\int\!dt\frac{d}{dt}\Theta\left(\vec x_i(t)-\vec
  x_j(t)\right)\right).
\eeq
Using this representation, the weighted 
path integral eq.~(\ref{mulconpi}) becomes
\bea\label{expsapi}
&&K(q^\prime,t^\prime;q,t)=\int_{q(t)=q;\>q(t^\prime)=q^\prime}\!\!\!\!
\!\!\!\!\!\!\!\!Dq(t_0)\>e^{i\int_t^{t^\prime}\!dt_0\,\left(L[q(t_0)]
-\sigma\sum_{i\not=j} {d\over dt_0}\Theta[\vec x_i(t_0)-\vec
  x_j(t_0)]\right)}\\ 
&&\>=\!\!\!\!\!\!\sum_{n_{ij,\>(i\not=j)}\,=-\infty}^\infty
e^{-i\sigma\left(\sum_{i\not =j}\hat\Theta\left(\vec x_i(t^\prime)-\vec
  x_j(t^\prime)\right)+2\pi n_{ij}\right)}
K_0^{(n)}(q^\prime,t^\prime; q,t)e^{i\sigma\sum_{i\not= j}\hat\Theta\left(\vec x_i(t)-\vec
  x_j(t)\right)}.\nonumber
\eea
Hence, the weights $\chi$ can be viewed as the consequence of having added
  to the Lagrangian $L$ the 
  effective interaction term 
\beq
\label{effl}
L_{\rm eff}[q(t)]=-\sigma\sum_{i\not=j} {d\over dt}\Theta[\vec x_i(t)-\vec
  x_j(t)].
\eeq
If the starting Lagrangian $L$ described bosonic exitations, the interaction
eq.~(\ref{effl}) endows these excitations with statistics
  $\sigma$. 

Equation~(\ref{expsapi}) shows that the effect of the
statistics-changing interaction can be absorbed in a  redefinition of
the wave function by a phase:
\beq
\label{pathwf}
\psi_0(q,t)=e^{i\sigma\sum_{i\not=j}\Theta_{ij}(t)}\psi(q,t):
\eeq
the wave function $\psi_0$ is propagated by the path--integral defined
in terms of the bosonic Lagrangian $L$. However, it is defined on a space
of paths rather than a space of coordinates, and it satisfies
``twisted'' boundary conditions: upon rotation by $2\pi$ it
acquires a phase 
\beq
\label{twistbc}
R^{2\pi} \psi_0(q,t)=e^{i2\pi \sigma n(n-1)}\psi_0(q,t),
\eeq
and correspondingly the spectrum of eigenvalue of the angular momentum
operator (which generates
rotations) is 
\beq
\label{jshift}
j=j_0+\sigma n(n-1), 
\eeq
where $j_0$ is the spectrum of angular momentum for the original
Lagrangian.
We see explicitly that for a system of of particles the angular
momentum of the pair is shifted by $2\sigma$ i.e. each particle has
acquired spin $\sigma$.

The effective statistics-changing  Lagrangian $L_{\rm eff}$ eq.~(\ref{effl})
looks intrinsically nonrelativistic, in that it depends on the polar
angle as a function of time. However, it also admits a covariant
formulation, which will turn out to be closely related to the
formulation of a path integral for spin.
To see this, define a
covariant particle current
\beq\label{covcurr}
j^\mu=
\sum_{i=1}^n\left(1,{d\vec x_i\over dt}\right)
\delta^{(2)}\!\left(\vec x- \vec x_i\right)
=\sum_{i=1}^n\int \!ds\,\delta^{(3)}\!\left(x- x_i\right)
{d x^\mu\over ds},
\eeq
where $s$ is any covariant parametrization along the particle path
(e.g. the path-length).
Furthermore, add to the action
 $I_0=\int dt L(t)$ a covariant coupling of the current
to a gauge potential
$A_\mu$:
\bea\label{act}
I&=&I_0+I_c+I_f\\
\label{curcoup}
I_c&=&\Int 3 x j^\mu(x)A_\mu(x)\\
\label{csterm}
I_f&=&-{1\over 2\sigma}
\Int 3 x \epsilon^{\mu\nu\rho}A_\mu(x)\partial_\nu A_\rho(x).
\eea

The action $I_c$ for the gauge potential $A_\mu$ is quadratic and can
be integrated out:
\beq
\label{topeffact}
I_{\rm eff}[j]\equiv-i\ln\int\! {\cal D}A^\mu\, e^{i(I_c+I_f)}=
\pi \sigma\int\!d^3x\,d^3y\,j^\mu(x)K_{\mu\nu}(x,y)
j^\nu(y),\eeq
in terms of the  Green function $K_{\mu\nu}(x,y)$ for the operator
$\epsilon_{\mu\rho\nu} \partial_\nu$:
\bea\label{kdef}
K_{\mu\nu}(x,y)&=&-{1\over2\pi}\epsilon_{\mu\rho\nu} {(x-y)^\rho\over
|x-y|^3}\\\label{kgreen}
\epsilon_{\mu\nu\rho}\partial_\nu
K^{\rho\sigma}(x,y)&=&\delta_\mu{}^\sigma 
\delta^{(3)}(x-y).
\eea

The effective current-current interaction 
\beq\label{effactdir}
I_{\rm eff}=\sigma\sum_{i,j}I_{ij},\quad
I_{ij}=-{1\over 2}\int \! dx_i^\mu\,dx_j^\nu \epsilon_{\mu\rho\nu}
{\left(x_i-x_j\right)^\rho\over|x_i-x_j|^3}
\eeq
is formally identical to the interaction of the current $j^\mu$ with a
Dirac magnetic monopole potential $\tilde A_\mu$: 
\beq
\label{dirmonpot}
{x^\mu\over|x|^3}=\epsilon^{\mu\alpha\beta}\partial_\alpha\tilde
  A_\beta(x).
\eeq
It is now easy to recover  the form eq.~(\ref{effl}) of the
  spin-statistics changing interaction. To this purpose, we choose an
  explicit  ``Coulomb
  gauge'' representation for the potential $\tilde A_\beta(x)$:
\beq\label{cgdirpot}
\tilde A_\mu(t,\vec x)=
\left(0,-{\epsilon_{ab}x^b\over
  r(t-r)}\right),\quad r^2=|x|^2=t^2-x_1^2-x_2^2,
\eeq
and we parametrize paths with time, $s=t$.

We get
\bea
I_{ij}&=&-{1\over 2}\int_0^T \!dt\int_0^T\!dt^\prime{dx_i^\mu(t)
\over dt}
\left(\partial_\mu\tilde A_\nu(x_i-x_j)-\partial_\nu\tilde A_\mu(x_i-x_j)
\right) {dx_j^\nu(t^\prime)\over dt^\prime}\nonumber\\
&=&\int_0^T\!dt\, \epsilon^{ab}\left({dx_i^a\over dt}-{dx_j^a\over dt}\right)
{(x_i(t)-x_j(t))^b\over
|x_i(t)-x_j(t)|^2}+ I_g,
\label{spinactz}
\eea
where $I_g$ is
a rotationally invariant surface term which has no effect on spin and
statistics.
Now, terms with
$i=j$ in eq.~(\ref{spinactz}) vanish by antisymmetry, while terms
with $i\not=j$ can be rewritten using the identity
\beq
\partial_a\Theta(\vec
  x)=-\epsilon^{ab}{x^b\over|x|^2},
\label{curlid}\eeq
which immediately implies that 
\beq
\label{neweqold}
I_{ij}=-\int\!dt\,{d\over dt} \Theta(\vec x_i-\vec x_j)+I_g,
\eeq
i.e., up to the irrelevant $I_g$, the same as the action obtained from
the effective Lagrangian eq.~(\ref{effl}).

Summarizing, we have found that nontrivial statistics is enforced by
weighing topologically inequivalent paths in the path integral, that
inequivalent paths are those which correspond to interchanging the
coordinates of two or more particles, and that these weights can be
obtained as the result of adding to the Lagrangian an effective
interaction term, which shifts the spectrum of the total angular momentum
by a half-integer contribution per
particle. Furthermore, in two dimensions we have obtained an explicit
local representation of this effective interaction term, which is
formally equivalent to the interaction of the particle current with a
Dirac magnetic monopole localized on each other particle.

\section{A Path Integral for Spin}
\label{sec:spinpi}
Spin is usually quantized by introducing degrees of freedom which live in
an internal space. In particular, the quantization of Fermions is
usually performed by introducing anticommuting variables.
However, in the previous section we have seen that it is possible to
represent the effect of fermionic statistics in terms of an
interaction defined in configuration space, and then path-integrating
over this space.
 In this
section we shall see that it is also possible to obtain the
path--integral quantization of a spin degree of freedom by
constructing the configuration space for a classical spin, and then
path-integrating over evolutions in this configuration spaces with a
suitable weight.

\subsection{The Spin Action}
\label{sec:spinact}
It is well-known that the classical action for a 
free (relativistic) particle coincides with the arc-length $L$ of the path $x^\mu(s)$
traversed by it, and in fact its quantization~\cite{feynhibbs} 
can be obtained by 
by summing over paths with a weight 
given by an action which coincides with the arc-length $L$:
\beq
\label{spinless}
 I=m\int ds \sqrt{\left(
\frac{d x^\mu}{ds}\right)^2}=m L.
\eeq
Hence, the quantization of the spinning particle is obtained by first
defining the space of paths, and then introducing as a weight over it
the simplest geometric invariant of the paths.

The path--integral quantization of spin can be obtained in a similar
way. The
configuration space for spin is the set of points spanned by a
vector 
\beq
\label{spinnorm}
\vec s= \sigma \vec e
\eeq
 with fixed modulus $|\vec s|=\sigma$, namely the two-sphere
$S^2$. This can be viewed as the result of the action of
the rotation group on a reference vector, namely, the coset of the
rotation group over the subgroup of rotations that leave the reference vector
invariant (little group): 
$S^2=SO(3)/SO(2)$. The simplest invariant over this space is the solid
angle subtended by a closed path. Therefore, parametrizing the vector
$\vec e$ in spherical coordinates
\beq\label{sphcoor}
\vec e=\pmatrix{\sin \theta  \cos \phi \cr
\sin \theta\sin\phi \cr\cos \theta \cr}
\eeq
we define the 
{ spin action} as
\beq\label{spinact}
I_s=\int\!dt\,{\cal L}(\theta,\phi)=s\int\!dt\,\cos\theta\dot\phi.
\eeq

Equation~(\ref{spinact}) provides us with a spin action in the sense
that the time--evolution (transition amplitude)  for spin wave functions 
\beq
\label{spinwf}
|\phi\rangle=|m\rangle\langle m|\phi\rangle;\quad \langle
m|\phi\rangle={e^{-im\phi}\over \sqrt{2\pi}}
\eeq
is given by
\beq
\label{spinpi}
\langle f | i \rangle=\langle \phi_f  |e^{i\int H(t)\, dt}|\phi_i
  \rangle,=
\int_{\vec e(t_f)=\vec e(\phi_f);\>
\vec e(t_i)=\vec e(\phi_i)}\!\!\!\!\!\!\!\!\!\!\! \!\!\!\!\!\!\!
D \vec e \>\>\>\>e^{i\int\!dt\,{\cal
L}_s-V(\vec J)},
\eeq
where $H(t)$ is a Hamiltonian which describes the spin  dynamics
  (e.g. the coupling with an external magnetic field, $H=\vec s\cdot
  \vec B$) and the   boundary conditions are given
 in terms of $\phi$ only (which is equivalent to specifying an
  eigenvalue $m$ of the third component of angular momentum).
This result can be proven by direct
  computation~\cite{nielsen,johnson}. We shall instead first show that
  the action $I_s$ eq.~(\ref{spinact}) leads to the correct classical
  dynamics of spin, then quantize it using the general principles of
  geometric quantization.

Let us first take a closer look at the spin action. Its geometric
interpretation becomes apparent by rewriting it as
\bea\label{actcur}
I_s&=&\sigma\int_C\cos\theta\dot\phi dt=\sigma\int_C\cos\theta d\phi\\ 
\label{actsur}&=&\sigma\int_S d\cos\theta d\phi=\sigma\int_Sd\vec S\cdot \vec
e=\sigma\int_S\left({ 
\partial \vec e\over
\partial s} \times{\partial \vec e\over
\partial t}\right)\cdot \vec e,\eea
where $C$ is the path travsersed by the vector $\vec e$
eq.~(\ref{sphcoor}), and the second step eq.~(\ref{actsur}) holds when
the path is closed, in which case 
$S$ is th surface bound by $C$.
In such case, eq.~(\ref{actsur}) shows explicitly that the 
spin action coincides with the solid angle subtended by $C$

Equation~(\ref{actsur}) shows manifestly that there is a $4\pi$
ambiguity in the definition of the spin action, which corresponds to
the possibility of choosing the upper or lower solid angle subtended
by C on the sphere. In order for this ambiguity to be irrelevant, the
parameter $\sigma$, which as we shall see corresponds to the value of
spin, must be quantized in half-integer units.
The connection between the spin action and the effective
two--dimensional statistics action eq.~(\ref{effactdir}) becomes clear
by rewriting it as
\beq
\label{newoldsa}
I_s=\sigma\int_S \!d\vec S\cdot\vec
\nabla \times \tilde{\vec  A}[\vec e],\eeq
where $\tilde A$ is the   Dirac monopole potential
eq.~(\ref{dirmonpot}), in the space of spin vectors:
\beq\label{dirmontpote}
\vec e=\vec \nabla \times \tilde{\vec A}[\vec e]\eeq

\subsection{Classical Dynamics}
\label{sec:cdyn}

In order to verify that the spin action defines the action for a
classical spin degree of freedom, we first check that it leads to the 
 Poisson bracket
\beq
\label{spinpbs}
\{s^i,s^j\}=\epsilon^{ijk}s^k.
\eeq
This is easily done using the
Faddeev-Jackiw formalism~\cite{fadjack} for 
the Hamiltonian treatment
of systems defined by first-order Lagrangians, i.e. by a Lagrangian of
the form
\beq\label{folag}
L=f_i(x){d x_i\over dt} -V(x).
\eeq
Namely, it easy to see that the Euler-Langrange equations implied by
the Langrangian eq.~(\ref{folag}) have the form
\bea\label{eleq}
&&f_{ij}{d x_j\over dt}={\partial V\over\partial x_i}\\\label{fdef}
&&\quad f_{ij}\equiv{\partial f_j\over \partial
	      x_i}-{\partial f_i\over \partial x_j}.
\eea
This coincides with the canonical Hamiltonian form
\beq
\label{hameq}
{d x^i\over dt}=\{x^j,x^i\}{\partial V\over \partial
  x^j}=\{V,x^i\}\eeq
if the Poisson brackets are given by
\beq\label{genpbs}
\{x^i,x^j\}=(f^{-1})^{ji}
\eeq
It can be shown that the same result is found in the  more standard approach, where the
  Lagrangian eq.~(\ref{folag}) is viewed as defining a constrained
  dynamics, which is then treated defining suitable
Dirac brackets.

Specializing this formalism to the spin action we see that  its  Dirac
monompole form eq.~(\ref{newoldsa}) has the form of
eq.~(\ref{folag}) with
\beq\label{fidef}
f_i=\sigma\tilde A_i[\vec e].
\eeq
Using the definition eq.~(\ref{fdef}) this leads to 
\beq\label{fspin}
f_{ij}=\sigma\left(\partial_i\tilde A_j-\partial_j\tilde A_i\right)
=\sigma\epsilon^{ijk}e^k.
\eeq
Because 
\beq\label{ffinv}
f_{ij}^{-1}={1\over \sigma^2}f_{ij}
\eeq
the Poisson Brackets are
\beq\label{espinpbs}
\{e^i,e^j\}={1\over \sigma}\epsilon^{ijk}e^k,
\eeq
which, identifying  the spin vectore $\vec s$ with
\beq\label{spinvec}
\vec s= \sigma
  \vec e,
\eeq
immediately lead to the  spin Poisson brackets
eq.~(\ref{spinpbs}). This also shows that the parameter $\sigma$ gives
the value
of spin.
\subsection{Geometric Quantization}
\label{sec:geoq}

The spin action can be quantized using the formalism of geometric or
``coadjoint orbit'' quantization~\cite{polyakov,wiegmann}. Namely,
we  view the time evolution of the (unit)
spin vector $\vec e(t)$ as the result of the action of a
rotation matrix $\Lambda(t)$ on a reference vector $\vec e_0$:
\beq\label{orbit}
\vec e(t)=\Lambda(t) \vec e_0.
\eeq
This defines a path (orbit) in $S^2=SO(3)/SO(2)$ , where $SO(2)$ 
is the little group of $\vec
e_0$ (the set of $\Lambda$ matrices which leaves $\vec e_0$ invariant).

The path in $S^2$ can be lifted to a path in 
$SO(3)$  by assigning a frame, e.g. by defining the vector
\beq\label{normal}
\vec n(t)\equiv \frac{ \dot {\vec e}(t)}{|\dot {\vec e}(t)|}
\eeq
which satisfies
\beq\label{normnorm}
{\vec n}\cdot\vec e=0.
\eeq
The triple $\vec e$, $\vec n$, and
\beq\label{bin}
\vec b(t)\equiv\vec e(t)\times \vec n(t).
\eeq
defines a time--dependent frame, which coincides with the standard
Frenet frame if $\vec e(t)$ is viewed as  the tangent vector to some
path, in which case $\vec n$ and $\vec b$ are the unit normal and
binormal, respectively. The matrix $\Lambda$ is then  fully specified
by eq.~(\ref{orbit}) and 
\beq\label{norel}
\vec n(t)=\Lambda(t) \vec n_0.
\eeq

It is convenient in particular to choose the set of reference vectors
\beq\label{refvecdef}
\pmatrix{\vec v^{(3)_0}=\vec
  e_0\cr \vec v^{(1)}_0=\vec n_0\cr \vec v^{(2)}_0=\vec b_0}
\eeq
as
\beq\label{refvecform}
{v^{(a)}_0}_i=\delta^a_i.
\eeq
It is then easy to see that
the quantity
\beq\label{mcdef}
\left(\Lambda^{-1}\dot\Lambda\right)^{ij}=
\vec v^{(i)}\cdot\dot{\vec v}^{(j)}
\eeq
is an element of the $SO(3)$ algebra, the so--called
Maurer-Cartan form, given by
\beq\label{mcexp}
\left(\Lambda^{-1}\dot\Lambda\right)_{ij}=\sum_{ab}
C_{ab}(M^{ab})_{ij};\quad
(M^{ab})_{ij}=\left(\delta_i^a\delta_j^b-\delta_j^a\delta_{i}^b\right).
\eeq
The coefficients $C_{ij}$ can be extracted by exploiting the fact that
the generators are orthogonal under tracing:
\beq
\label{ctrace}
C_{ij}={1\over4}\tr\left(M_{ij}\Lambda^{-1}\dot\Lambda\right)={1\over2}
\vec v^{(i)}\cdot\dot{\vec v}^{(j)}.\eeq

We can now use this gometric formalism to rewrite yet again the spin
action eq.~(\ref{spinact}) as
\bea
I_s&=&\sigma\int_S\left({
\partial \vec e\over
\partial s} \times{\partial \vec e\over
\partial t}\right)\cdot \vec e=\sigma\int \!dt\,\,\dot{\vec b}\cdot\vec
n+\hbox{\rm integers}\nonumber
\\\label{spingeom}
&=&\sigma\left(\tr\int\!dt\,{1\over2}
\left(\Lambda^{-1}\dot\Lambda M_{12}\right)+\hbox{\rm integers}\right).
\eea
Note that any spin-dependent potential $V(\vec \sigma)$ can be re-written in terms of
$\Lambda$ by exploiting eq.~(\ref{ctrace}) to express the spin vector
$\vec e$ in terms of
$\Lambda$ :
\beq
\label{spintolam}
e^i=\sigma\epsilon^{ijk}\left(
\Lambda^{-1}{M_{12}\over2}\Lambda\right)_{jk}.\eeq

This new form eq.~(\ref{spingeom}) of the spin action has a twofold
advantage: first, it does not depend on the representation, and
second, it is amenable to geometric quantization. To demonstrate its
representation-independence, let us show how the spinor
representation is recovered from it. For spin $\half$, the generators are
\beq\label{paulmat}
M_{ij}=-i\epsilon^{ijk}\sigma_k,
\eeq
where $\sigma_i$ are the usual
 Pauli matrices.
We then have
\beq\label{sphalfact}
\tr{1\over2}\left(\Lambda^{-1}\dot\Lambda M_{12}\right)=\tr\left(
\Lambda^{-1}\dot\Lambda{\sigma_3\over 2i}\right)=\tr\left(
\Lambda^{-1}\dot\Lambda\left({\1+\sigma_3\over2i}\right)\right).
\eeq

The connection to (Pauli) spinors is found by introducing the
reference two--component spinor
\beq\label{refspinor}
\psi_0=\pmatrix{1\cr0\cr},
\eeq
upon which the matrix $\Lambda$ is taken to act in the spinor
representation, namely
\beq\label{spinorpath}
\psi(t)=T[\Lambda(t)]\psi_0,
\eeq
where $T[\Lambda(t)]$ is the spinor representation of the rotation
$\Lambda$.
The relation between the spinor and vector representation is provided
by constructing spinor bilinears
\beq
\label{lamtovec}
\psi^*\sigma^i\psi=\Lambda_{ij}
\psi_0^*\sigma^i\psi_0=\Lambda_{ij} e_0^j.
\eeq
Using this relation, and noting that
\beq\label{refproj}
|\psi_0\rangle\langle\psi_0|=\left({\1+\sigma_3\over2}\right)
\eeq
it is easy to rewrite the spin action $I_s$
eq.~(\ref{spingeom}) as
\beq\label{paulspinact}
I_s={1\over2}\int\!{dt\over i}\,\psi^*(t){d\over
  dt}\psi(t),
\eeq
which has the form of the kinetic term for a Pauli spinor $\psi(t)$.
A generic spin--dependent  potential $V(\vec\sigma)$ can be written in terms of $\psi$
by using the relation
\beq\label{etopsi}
\vec e
=\psi^*(t)\vec \sigma\psi(t).
\eeq

The quantization of spin is now reduced to the general
problem~\cite{wiegmann} 
of 
quantizing a system whose confiuration space is the space of 
states $|\psi\rangle$  which are orbits of a group
$G$:
\beq\label{torbit}
|\phi\rangle=T(g)|\phi_0
\rangle,\eeq
 where $g\in G$ is an element of the group of which $T(g)$ provides a
 unitary representation. The axioms of quantum mechanics imply
 that transition amplitudes for this system are given
 by the path integral
\beq\label{pigq}
\langle f|i\rangle=\int\! Dg\, e^{iI_w[g]}
\eeq
with the action
\beq\label{actgq}
I_w[g]=\int\!dt\,\langle\phi_0|\left[
T(g^{-1}(t)){d\over i dt}T(g(t))-H(g(t))\right]|\phi_0\rangle,
\eeq
where $H$ is a generic spin--dependent potential (or Hamiltonian,
which coincides with the potential for a first-order Lagrangian).

The spin action eq.~(\ref{spingeom}) is seen to coincide with the kinetic term of the geometric
action eq.~(\ref{actgq}) if one identifies
the representation matrix $T(g)$ with $\Lambda$ eq.~(\ref{orbit}), and
one observes that the projector on the state $|\phi_0\rangle$ can be
expressed in terms of the generator
$C^0_{ij}M^{ij}$ of
the little group of $|\phi_0\rangle$:
\beq\label{projgen}
|\phi_0\rangle\langle\phi_0|=C^0_{ij}M^{ij}.
\eeq
Indeed, we get
\beq\label{spinfin}
\int\!dt\,\langle\phi_0|\left[
T(g^{-1}(t)){d\over i dt}T(g(t))\right]|\phi_0\rangle=
\int\!dt\, \tr C^0_{ij}M^{ij} \Lambda^{-1}(t)\dot \Lambda(t)
\eeq
which coincides with the spin action if we choose
$C^0_{ij}M^{ij}=M_{12}$.
Hence, the spin path integral eq.~(\ref{spinpi}) with  the spin action
eq.~(\ref{spinact}) 
follows from geometric quantization of the space of $SO(3)$ orbits.

The relation of this result to the usual sum over paths \`a la
Feynman is apparent if we specialize again
to the case of spin
 $\half$.
The sum over paths is performed by dividing the time evolution from $t_i$
to $t_f$ into discrete time steps $\Delta t={t_f-t_i\over N}$ so that
$t_j=t_i+(j-1)\Delta t$, and then letting $N\to\infty$.
For a spin system we get
\beq\label{feynsum}
\langle f| i\rangle=\langle\psi_f|e^{-i\int_{t_i}^{t_f}\! H \,dt}|\psi_i\rangle
=\prod_{j=1}^N\int\!d\Lambda_j
\langle\psi_{j+1}|e^{-i\Delta t H(t_j) }|\psi_j\rangle.\eeq
The evolution along an
infinitesimal time slice is then given by
\bea
\langle\psi_{j+1}|e^{-i\Delta t H(t_i) }|\psi_j\rangle
&\approx& 
\langle\psi_{j+1}|\left(1-i\Delta t
H(t_j)\right)|\psi_j\rangle\nonumber\\
\label{inftevol}
&=&1+{1\over2}\Delta t \psi^*{d\over dt}\psi -i\Delta t
H(t_j)\\&\approx& 
e^{i\left[\psi^*{d\over i dt}\psi- \Delta t H(t_i)\right]},\nonumber\eea
which coincides with  the geometric quantization result
eq.~(\ref{actgq}). The first-order quantization of spin is a simple
consequence of the fact that a spin Hamiltonian does not contain a
quadratic kinetic term: the action is then entirely determined by the
first-order parallel transport of the spin vector.

The meaning of these results is that first, the probabilty for the
time evolution between two spin states is given by
\beq\label{spinevolfin}
\langle f| i\rangle=
\int_{\vec e_f(t_f)= \vec e_f(\Lambda_f);\>
\vec e_i(t_i)= \vec e_i(\Lambda_i)
}\!\!\!\!\!\!\!\!\!\!\!\!\!\!\!\!\!\!\!\!\!
D \vec e\>\>\>\>\>\> e^{i\left[I_s[e]-\int\!dt\,H(t,\vec e)\right]} 
\eeq
and furthermore, the matrix element of any spin--dependent operator $
F(\vec \sigma)$ can be determined as
\beq\label{matelevol}
\langle f|  F(\vec \sigma)|i\rangle=
\int_{\vec e_f(t_f)= \vec e_f(\Lambda_f);\>
\vec e_i(t_i)= \vec e_i(\Lambda_i)
}\!\!\!\!\!\!\!\!\!\!\!\!\!\!\!\!\!\!\!\!\!
D \vec e\>\>\>\>\>\> e^{i\left[I_s[e]-\int\!dt\,H(t,\vec e)\right]}  F(\vec e).
\eeq

Summarizing, we have seen that the path-integral quantization of a
``static'' spin degree of freedom --- as e.g. in the
Heisenberg model --- can be given in terms of a geometrically
determined first--order spin action. The usual formalism in the
spin-$\half$ case is obtained by specializing to the spinor
representation of spin vectors, but it does not require anticommuting
variables or relativity. It is interesting to note that the same results can be
obtained from the well-known ``Schwinger boson'' 
representation of angular momentum
operators  in terms of creation and
annihilation operators for the (bosonic) harmonic
oscillator~\cite{schwinger}, by quantizing the harmonic
oscillator degrees of freedom in terms of a first-order action~\cite{wiegmann}
(i.e. in terms of coeherent states).

\section{Relativistic Spinning Particles}
\label{sec:relpart}
As discussed in the introduction, in a relativistic theory physical
states are irreducile representations of the Poincar\'e group,
i.e. they carry mass and spin: the 
one-particle state $|m,s\rangle$ satisfies
\beq
\label{opartst}
P^2 |m,s\rangle= m^2 |m,s\rangle;\quad W^2|m,s\rangle=m^2 s(s+1),
\eeq
where the Pauli-Lubanski operator $W$, defined in 
eq.~(\ref{paluop}), generates Lorentz tranformations which
leave the particle momentum invariant, because  by construction
$W_\mu P^\mu=0$. In particluar, in the rest frame of the particle  (for massive particles)
  $p=(m,\vec 0)$, so $W=(0,\vec s)$. In a general frame,
spin spans the three dimensional ($d-1$ dimensional) space
  orthogonal to momentum.
 This
introduces a coupling between spin and momentum which determines the
dynamics of a relativistic spinning particle, both at the classical
and quantum level.

\subsection{Path Integral for Spinless Particles}
\label{sec:pispinl}
Before discussing the quantization of spinning particles, let us
review the path--integral quantization of a massive spinless
particle~\cite{feynhibbs}. 
As we mentioned already, the action eq.~(\ref{spinless}) of a spinless free particle, or
the kinetic term in the action for an interacting spinless particle,
coincides with the arc-length of the path traversed by the
particle. This can be written in various equivalent ways: 
the simple integral of the arc-length element $ds=\sqrt{dx^\mu dx_\mu}$
eq.~(\ref{spinless})
can be rewritten in terms of an induced metric $g(s)$ along the path
\beq\label{spinlessind}
I_0=\int ds\left[\frac{1}{\sqrt{g}} \half \left(
\frac{d x^\mu}{ds}\right)^2+\frac{m^2}{2}\sqrt g\right].
\eeq
Both at the classical and at the quantum level, the equation of
motion for $g$ is the constraint
\beq\label{geq}
g=\frac{\dot x^2}{m^2}, 
\eeq
which shows that indeed $g(s)$ is the induced metric
\beq\label{indmet}
dx^2= g(s) ds^2,
\eeq
and leads back to the original form eq.~(\ref{spinless}) of the action
when substituted in eq.~(\ref{spinlessind}).

The action eq.~(\ref{spinlessind}) can in turn be rewritten in first-order form 
\beq\label{spinlessfo}
I_0=\int ds \left[p_\mu \frac{dx^\mu}{dt}-\frac{\sqrt
  g}{2}\left(p^2-m^2\right)\right],
\eeq 
where the momentum (tangent vector) $p^\mu$ is also fixed by a
constraint
\beq\label{momconstr}
p^\mu=\frac{1}{\sqrt g}\dot x^\mu
\eeq
which again leads back to the original form eq.~(\ref{spinlessind})
when subsitituted in the action eq.~(\ref{spinlessfo}). This
first--order form of the action is the most suitable for geometric
quantization, i.e. for describing the dynamics of the spinning
particle similarly to the way we have described the dynamics of spin
in section~\ref{sec:geoq}.
The classical equations of motion can be obtained from any of these
equivalent forms of the action, and express  energy-momentum
conservation. For instance, using the first-order form
eq.~(\ref{spinlessfo}) we get immediately
the Euler-Lagrange equations
\beq\label{clelag}
\frac{d}{dt} p^\mu=0,\quad p^2=m^2.\eeq

Path--integral quantization~\cite{feynhibbs} can be perfomed by
exploiting the ``gauge invariance'', i.e. the reparametrization 
invariance of the system~\cite{polyakov}. The 
(Euclidean) path integral 
\beq\label{eucpi}
\langle x^\prime|x\rangle={\cal N}\int_{x(0)=x;\>x(1)=x^\prime}\!\!\!\!\!\!\!\!
\!\!\!\!\!\!\!\!\!\!\!Dx(s)\>\>e^{-m\int_0^1\!ds\,\sqrt{\dot x^2}}
\eeq
can be rewritten introducing the
induced metric $g(s)$ eq.~(\ref{geq}) as
\beq\label{piind}
{\langle x^\prime|x\rangle={\cal N}\int_{x(0)=x;\>x(1)=x^\prime}\!\!\!\!\!\!\!\!
\!\!\!\!\!\!\!\!\!\!\!Dx(s)Dg(s)\>\>\delta^{(\infty)}\! \left(\dot x^2-g\right)
e^{-m\int_0^1\!ds\,\sqrt{g}}.}
\eeq
Reparametrization invariance is now manifest, because upon a general reparametrization
$s\to f(s)$, the metric $g(s)$ transforms as
$g(s)\to g(f(s))[\dot f(s)]^2$.
We can now perform the path integral by fixing the gauge, e.g. by imposing the condition
\beq\label{gauge}
\dot g(s)=0.
\eeq
Because  the  path-length is
\beq
\label{pl}L=\int_0^1\!ds\, \sqrt{\dot x^2}
=\int_0^1\!ds\, \sqrt{g(s)}
\eeq
the gauge condition~(\ref{gauge})
implies
\beq\label{gtol}
g(s)=
L^2.
\eeq
We can thus write the gauge-fixed path-integral as
\bea\label{gfpi}
\langle x^\prime|x\rangle&=&{\cal N}\int_0^\infty\! dL\,\int_{x(0)=x;\>x(1)=x^\prime}
\!\!\!\!
\!\!\!\!
\!\!\!\!\!\!\!\!\!\!\!Dx(s)Dg(s)\>\>\delta^{(\infty)}\!
\left(\dot x^2-g\right)
\delta\! \left(g-L^2\right)  
e^{-mL}\nonumber\\
&=&{\cal N}\int_0^\infty\! dL\,\int_{x(0)=x;\>x(1)=x^\prime}
\!\!\!\!
\!\!\!\!
\!\!\!\!\!\!\!\!\!\!\!Dx(s)\>\>\delta^{(\infty)} \left(\dot x^2-L^2\right)
e^{-mL}.\eea
After gauge-fixing, a residual integration over path lengths $L$ remains.

The path-integral can be re-written in terms of geometric variables
along the path: this leads to geometric quantization again. 
We introduce  a tangent vector
along the path, which for classical paths (those which satisfy the
Euler-Lagrange equations) coincides with  the particle four-momentum:  
\beq\label{clmom}
e^\mu={\dot x^\mu\over
  |\dot x|}={\dot x^\mu\over L}.
\eeq
We can replace the path-integration over trajectories by a
path-integration over the tangent vectors $e^\mu$. However, the
boundary conditions now become a non-local constraint:
\beq\label{bcconst}
{x^\mu}^\prime-x^\mu=\int_0^L\!ds\, e^\mu(s)).
\eeq
We thus get finally
\bea\label{finpi}
\langle x^\prime|x\rangle&=&{\cal N}\int_0^\infty\! dL\,\int
De(s)\>e^{-mL}
\delta^{(\infty)}\! \left( e^2-1\right)
\delta^{(3)}({x^\mu}^\prime-x^\mu-\int_0^L\!ds\, e^\mu(s))\nonumber \\
&=&{\cal N}\int\! dL\,d\vec p\int
De(s)\>e^{-mL}
\delta^{(\infty)}\! \left( e^2-1\right)
e^{i p\cdot \left(x^\prime-x-\int_0^L\!ds\, e(s)\right)}.
\eea

The usual expression of the bosonic (Klein-Gordon) propagator is
obtained by regularizing the formal expression eq.~(\ref{finpi}). To
this purpose, we cut off paths
which are coarse on a scale $\sim \epsilon$ (where, of course
$\epsilon$ has the dimensions of
[length]). We then take the continuum limit with a mass
renormalization condition, expressed by defining a renormalized
mass $M_{\rm phys}$ such that
\beq\label{rencond}
m\propto\varepsilon M_{\rm phys}^2.
\eeq
The propagator $K(p)$ is obtained as
the Fourier transform of the renormalized position--space amplitude:
\bea\label{kgprop}
K(p)&=&\lim_{\varepsilon\to0}{\cal N}\int
dL\,e^{-mL}\int De(s)e^{-\frac{\varepsilon}{2}\int_0^L\!ds\,\dot e^2
}e^{-i p\cdot \int_0^L\!ds\, e(s)}
\delta^{(\infty)}\! \left( e^2-1\right)\nonumber\\&=&{\cal N}
\int
dL\,e^{-L\varepsilon M_{\rm phys}^2} e^{ -L \varepsilon p^2}
={\cal N}\frac{1}{p^2+M_{\rm phys}^2}.
\eea
Up to the
irrelevant albeit infinite normalization
constant ${\cal N}$, we have thus recovered the standard form of the
Klein-Gordon propagator.

\subsection{The Classical Spinning Particle}
\label{sec:clspinp}

The spinning particle is now obtained by coupling a spin degree of
freedom
to the spinless
particle of section~\ref{sec:pispinl}, with
dynamics governed by the action discussed in section~\ref{sec:spinact}.
This can be done in an elegant geometric way by combining the
translational and spin configuration spaces. To this purpose, 
in one time and $d-1$
space dimensions, we define a set of $d-1$ orthonormal vectors $e^\mu$,
$n_{(1)}^\mu$,\dots,$n_{(d-2)}^\mu$, which can in turn be obtained by action
of a Lorentz transformation matrix $\Lambda$ on a set of reference vectors
\beq\label{jointconf}
\left\{\matrix{e^\mu&=\Lambda^\mu{}_\nu \hat t^\nu \cr {n_{(i)}}^\mu&=\Lambda^\mu{}_\nu 
\hat n_{0\,(i)}{}^\mu
\cr}
\right. .
\eeq
The reference vectors
\beq\label{basvect}
\hat t^\mu=\pmatrix{1\cr\vec 0\cr},\qquad
\hat n_0^{(i)}{}^\mu=\delta_i{}^\mu
\eeq define a basis
 in  one time and
$d-1$ space dimensions. The set of vectors $e^\mu$, $n_{(i)}^\mu$
 completely specifies the matrix $\Lambda$: indeed, the first vector
 has $d-1$ independent components (being unimodular), the second,
 orthogonal to it, has $d-2$ independent components and so on, so that
 overall they have $\sum_{i=0}^{d-2}(d-i-1)=\half d(d-1)$ independent
 components, like  the $O(d-1,1)$ matrix $\Lambda$. 

In the four-dimensional case we are interested in, the matrix
$\Lambda$ has  six independent components.
We take the vector $e^\mu$ as the unit tangent to the particle
trajectory, so that classically
is is identified with  momentum up to an overall factor of
$m$:
\beq\label{pfrome}
e^\mu={\dot x^\mu\over
  |\dot x|};\quad p^\mu= m e^\mu
\eeq
and at the quantum level it is the variable one
path-integrates over (compare eq.~(\ref{finpi})). 
The vector $n_{(1)}^\mu$ is then identified with the spin vector
discussed in the previous section, it has two independent components
and lives in the $S^2$ orthogonal to $e^\mu$:
\beq\label{sfromn}
e_\mu n_{(1)}^\mu=0;\quad s^\mu= \sigma n_{(1)}^\mu
\eeq
At the quantum level, the two independent vectors $p^\mu$ and $s^\mu$
entirely specify
the configuration of the system, whereas at the classical level the
canonical coordinate $x^\mu$ must also be given.

The action for the spinning particle is now simply obtained by
combining the action for the spinless particle eq.~(\ref{spinlessfo}) with the
spin action eq.~(\ref{spingeom}): by writing both in terms of
$\Lambda$, the momentum-spin orthogonality constraint is automatically
enforced.
We get
\beq\label{spinpartac}
  I=\int ds \left[p_\mu \frac{dx^\mu}{dt}-\frac{\sqrt
  g}{2}\left(p^2-m^2\right)\right]+\sigma\tr \left(\Lambda^{-1} \dot \Lambda M_{12}
  \right).
\eeq

It is straightforward to check that, at the classical level, the correct
dynamics is obtained: the Euler-Lagrange equations are found by
varying the action upon the most general Poincar\'e transformation, namely 
a translation of $x^\mu$, and a Lorentz
transformation of $\Lambda$. The variation upon translations gives
trivially the spinless equation of motion eq.~(\ref{clelag}) (energy-momentum
conservation). The most general Lorentz variation is
\beq\label{lorvar}
 \delta\Lambda=i\omega^{\mu\nu}
M_{\mu\nu} \Lambda,
\eeq
upon which the action transforms as
\bea\label{lorvaract}
\delta I &=&-i
\tr \left(\omega^{\mu\nu}M_{\mu\nu} K\right)+i\sigma\tr
\left(S{d\over dt} \omega^{\mu\nu}M_{\mu\nu}\right)\\ \label{kmunudef}
\quad &&K_{\mu\nu}\equiv\left(\dot x_\mu p_\nu - x_\nu \dot
p^\nu\right)\\
\label{smunudef}
&&S_{\mu\nu}=\sigma\left(\Lambda^{-1}M_{12}\Lambda\right)_{\mu\nu}.
\eea
Demanding that the action be stationary  leads to the Euler-Lagrange equations
\beq\label{eulagspinpart}
{d\over dt}\left(x^\mu p^\nu-x^\nu p^\mu + S^{\mu\nu}\right)=0.
\eeq
Equation~(\ref{eulagspinpart})  expresses the set of conservation laws
of a Lorentz invariant Lagrangian: in particular, the 
$(i,j)$ components give  the
conservation of (total) angular momentum, while the
$(0,i)$ components give the equation
$\vec p=\frac{d}{dt}\left( \vec x E\right)$
which relates momentum to velocity in the usual way.

\subsection{Quantum Spinning Particles and Fermions}
\label{sec:spinfer}
The dynamics of the spinning particle, described
by the action eq.~(\ref{spinpartac}), is given on the space of
Lorentz orbits $\Lambda(t)$ which evolve according to eq.~(\ref{jointconf}) the
pair of vectors $p^\mu$ eq.~(\ref{pfrome}),
$s^\mu$ eq.~(\ref{sfromn}). The path integral then follows from
geometric quantization~\cite{wiegmann,orland} eqs.~(\ref{pigq},\ref{actgq}):
\beq\label{pispinpart}
\langle x^\prime,\vec s^\prime|x,\vec s\rangle=\int\! d\vec p\,e^{i p\cdot \left(x^\prime-x\right)}\int
dL\,e^{-mL}\int D\Lambda(s) e^{-i  \int_0^L\!ds\,
 \left[ p\cdot\Lambda\hat t- \sigma\tr \left(\Lambda^{-1} \dot \Lambda M_{12}
  \right)\right]} .
\eeq
In practice, the path integral is found by combining the spin path
integral eq.~(\ref{spinevolfin}) and the spinless particle path integral
eq.~(\ref{finpi}).

Let us now discuss in particular the spin-$\half$ case in the spinor
formulation, and show how the Dirac equation is recovered.
We can do this  promoting to the Lorentz group the connection between
spinor and vector representations  of the rotation group
eq.~(\ref{etopsi}). This is based on
the transformation law of Dirac matrices, which connect
the four-vector representation $\Lambda$  of the Lorentz group with
the corresponding spinor representation $T(\Lambda)$:
\beq\label{dirgamtransf}
T(\Lambda^{-1})\gamma^\mu
  T(\Lambda)=\Lambda^\mu{}_\nu\gamma^\nu.
\eeq
Now, it is easy to show that  given an unimodular vector $v^\mu$, 
the spinor $\psi$ such that
\beq\label{psitoe} 
\psi^*
  \gamma^\mu \psi=v^\mu
\eeq
 satisfies  the condition
\beq\label{gendireq}
v_\mu\gamma^\mu\psi=\psi
\eeq
(in Euclidean space, in Minkowski space the spinor $\psi^*$ must be
replaced by $\bar\psi\equiv \psi^*\gamma^0$).

In our case, we associate to the one-particle state with normalized
momentum $e^\mu$ the spinor  $\psi[e^\mu]$  which satisfies the condition
\beq\label{truedireq}
p_\mu\gamma^\mu \psi= m\psi,
\eeq
i.e. the Dirac equation. In practice, we can determine $\psi[e^\mu]$
by acting with the spinor
representation $T(\Lambda)$ of the transformation $\Lambda$
eq.~(\ref{jointconf})
\beq\label{spintransf}\psi=T(\Lambda)\psi_0
\eeq
on the reference spinor $\psi_0$ such that
$\psi^*_0\gamma^\mu\psi_0=\hat t^\mu$, i.e. (using
eq.~(\ref{gendireq})
such that
\beq\label{zerdireq}
\gamma^0\psi_0=\psi_0.
\eeq
If one uses the so-called Dirac representation for the $\gamma$ matrices,
$\gamma^0=\pmatrix{\1& 0\cr0 & -\1\cr}$ (where each entry is a
$2\times2$ block), so
\beq\label{psizsol}
\psi=\pmatrix
  {\phi\cr
0\cr},
\eeq
where $\phi$ is any two--component spinor.

The condition that the spin vector be given by $s^\mu$ fixes entirely
the spinor (up to an overall U(1) phase): if $\Lambda$ is such that 
$\Lambda^{\mu}_\nu s^\nu_0= s^\mu$, then, choosing according to eq.~(\ref{basvect})
$s^\nu_0=\pmatrix{0\cr1\cr0\cr0\cr} $, the spinor $\psi$ is given by
\beq\label{spinspin}
\psi=T(\Lambda)\psi_0;\qquad 
\phi_0\equiv\pmatrix{1\cr0\cr}.
\eeq
It is easy to see that the spinor constructed in this way is an
eigenstate of the projection of the Pauli-Lubanski operator along the
spin vector $s^\mu=\half n^\mu$:
\beq\label{spinevec}
W^\mu n_\mu \psi=\pm \frac{m}{2}\psi(p,s).
\eeq
This is obvious in the rest frame, because then
 $W^\mu
  s_\mu=m s_i \epsilon^{ijk}\sigma_{jk}=m \vec s\cdot\vec\sigma$,
  where $\vec\sigma$ are Pauli matrices, and eq.~(\ref{spinspin})
  together with the relation between the spin vector and Pauli
  matrices eq.~(\ref{etopsi})
  implies that 
\beq\label{spineveccond}
\vec s\cdot\vec \sigma \phi=\pm \half \phi.
\eeq
In other words, in the rest frame $W^\mu s_\mu$ is just the standard
spin operator and $\phi$ is the
two-component spinor eq.~(\ref{spinorpath}) discussed in
sec.~\ref{sec:geoq}. But since in the rest frame the low components of
$\psi$ eq.~(\ref{psizsol}) vanish, this implies
\beq\label{weveccond}
W^\mu
  s_\mu\psi =\pm \half m \psi .
\eeq
In a generic frame, the four-vector $s^\mu$ is
boosted by $\Lambda$, so
\beq\label{nboost}
W^\mu n_\mu =W^\mu
  {\Lambda^{-1}}^{\nu}{}_\mu n_\mu=T(\Lambda) W^\mu
  T^{-1}(\Lambda)n_\mu,
\eeq
but  so is the spinor $\psi$
in such a way that [eq.~(\ref{spintransf})]
the eigenvector condition still holds:
\beq\label{boostcond}
T(\Lambda) W^\mu T^{-1}(\Lambda)n_\mu
T(\Lambda)\psi0=\pm \frac{m}{2} T(\Lambda)\psi_0
\eeq

Let us now consider the propagator $K(p)$, i.e. momentum--space path
integral, related by  Fourier transformation to
the path--integral eq.~(\ref{pispinpart}). We have found that
 in the spin-$\half$ case, if the spinor representation is
adopted, states along the path are instantaneous eigenstates of
$e_\mu\gamma^\mu$,
according to eq.~(\ref{truedireq}). In follows that
momentum eigenstates, which are the boundary conditions to the
momentum--space path--integral (i.e. states of definite $e^\mu$)
 automatically satisfy the Dirac equation. Furthermore, 
the spinor states satisfy
\beq\label{eeveccond}
\psi^*\gamma^\mu\psi=e^\mu,
\eeq
i.e., $e^\mu$ is obtained by acting with $\gamma^\mu$ on the
instantaneous spinor states along the path. But in
sect.~\ref{sec:geoq} we have proven [eq.~(\ref{matelevol})] that
the expectation value of any function $F(\sigma)$ can be obtained by
path--integration of the function $F(\Lambda)$ with a weight given by
the spin action itself. Applying this in
reverse, we see that averaging with the spin action produces the same
result as taking matrix element of instantaneous (path--otrdered)
functions of $\gamma^\mu$, where $\gamma^\mu$ is identified with
$e^\mu$ thanks to eq.~(\ref{eeveccond}).

The propagator is therefore given by
\bea
K(p)&=&\int
dL\,e^{-mL}\int D\Lambda(s) e^{-i \int_0^L\!ds\,
 \left[ p_\mu e^\mu- \sigma\tr \left(\Lambda \dot \Lambda M_{12}
  \right)\right]}\nonumber\\\label{ferprop}
&=& \int
dL\,e^{-mL}e^{-iL p_\mu \gamma^\mu}
\\
&=&\frac{1}{\thru p_\mu+m},
\nonumber\eea
i.e. the usual Dirac form.

The link with Fermi statistics is understood by observing that
the spin factor upon $2\pi$ rotation transforms as
\beq
\tr\left(\Lambda^{-1}\dot\Lambda R M_{12}R^{-1}\right)=R^i_j\hat
z^j\epsilon^{ijk} \tr\left(\Lambda^{-1}\dot\Lambda M_{jk}\right)
\eeq
so if $\sigma=\half$ the path--integral eq.~(\ref{pispinpart})
acquires a phase $e^{i\pi}=-1$.
In the more conventional approach, this follows from the anticommuting
properties of the $\gamma$ matrices, and it requires
anticommuting (Grassmann) variables. In the geometric approach which
we have followed this is not necessary, because the anticommutation
properties follows automatically from the fact that physical states
are localized on paths (so ordering along the path is enforced), and
paths are given weights that transform nontrivially upon
rotations. 
This provides an explicit realization of the general spin--statistics
relation derived in section~\ref{sec:spinstat}: once spin is obtained
as a consequence of an interaction defined in configuration space, the
link with statistics follows from the fact that particle interchange
can be perfomed by  $2\pi$ rotation.

Finally, it is interesting to observe that the dynamical coupling of
spin and momentum which follows from the geometric interpretation of
spin as a vector in the space which is orthogonal to momentum actually
changes the nature of the sum over paths: the
Hausdorff dimension of paths $d_h$ that contribute to the regularized
and renormalized Euclidean path integral in the continuum limit
is not the same for Bose and Fermi particles~\cite{jaro}. The
Hausdorff dimension  relates the typical length scale $L$ of paths which dominate the
propagator in the continuum limit to the momentum $p$ which is propagated:
\beq\label{haudef}
 L\sim p^{d_H}
\eeq
It can be proven  that $d_H=2$ for Bosons while  $d_H=1$ for Fermions~\cite{jaro}. A
rough and ready way to see this is to 
compare the bosonic propagator eq.~(\ref{kgprop}) and the fermionic  propagator eq.~(\ref{ferprop}):
it appears that the  scaling limit requires taking $L
m^a\sim$~constant
with $a=2$ for Bosons and $a=1$ for Fermions. 
This means that Bosonic paths are coarser then Femionc paths:
Bosonic propagation is an ordinary random walk (like Brownian motion), whereas Fermionic
propagation is a directed random walk, essentially because the spin
interaction quenches fluctuations of the tangent vector to the path.
\section{Conclusion}
The discussion of spin presented in these lectures 
was rooted in quantum mechanics, and
has used few field--theoretic
concepts. Yet, we have been able to derive many results which usually
require the full framework of relativistic quantum field theory: the
spin--statistics connection, multivalued spin wave functions, the spin
propagator, the Dirac equation.  In fact, we have shown that the
quantization of spin both in a nonrelativistic and a relativistic
setting follows from general properties of the configuration space for
orbits of the rotation group, viewed as a subgroup of the Galilei or
Poincar\'e group, respctively. It thus appears that the standard
field-theoretic approach is is
merely a convenient way of achieving the quantization of systems of
elementary excitations which provide irreducible representations of
the Galilei or Poincar\'e group, because field theory automatically 
combines quantum
mechanics with the relevant symmetry group in
a local, unitary way. Of course, the standard field--theoretic
approach, with anticommuting variables and spinors, is by far more
convenient for the sake of practical computations. However, we have
attempted to show that
the origin of the quantum field theoretic features of
spin in the way symmetry is realized in quantum mechanics.

{\bf Acknowledgement:} I thank Walter P\"otz 
for giving me the opportunity to present these  lectures in a
stimulating interdisciplinary environment, and all those who attended
the lectures for their interest and enthousiasm. I thank W.~Siegel for
pointing out the relevance of Ref.~\cite{schwinger}.

\index{paragraph}
%
%

%
%
%

%
%



\printindex
\end{document}